\documentclass[preprint]{iucr}
     \paperprodcode{a000000}      
     \paperref{xx9999}            
     \papertype{FA}               
     \paperlang{english}          
     \journalcode{A}
     \journalyr{2009}
     \journaliss{1}
     \journalvol{63}
     \journalfirstpage{000}
     \journallastpage{000}
     \journalreceived{0 XXXXXXX 0000}
     \journalaccepted{0 XXXXXXX 0000}
     \journalonline{0 XXXXXXX 0000}

\usepackage{amsmath,amssymb}
\usepackage{graphicx}
\usepackage{dcolumn}
\usepackage{bm}

\begin{document}

\title{Point processes for decagonal quasiperiodic tilings}
\shorttitle{Point processes for decagonal tilings}

\cauthor{Nobuhisa}{Fujita}{nobuhisa@tagen.tohoku.ac.jp}{}
\aff{Institute of Multidisciplinary Research for Advanced Materials, Tohoku University,
Sendai 980-8577 \country{Japan}}
\shortauthor{Fujita}

\maketitle                        

\begin{synopsis}
A general construction principle of inflation rules for decagonal
quasiperiodic tilings is proposed.
\end{synopsis}

\begin{abstract}
A general construction principle of inflation rules for decagonal
quasiperiodic tilings is proposed. The prototiles are confined to
be polygons with unit edges. An inflation rule for a tiling
is the combination of an expansion and a division of the tiles,
where the expanded tiles can be divided arbitrarily as far as the
set of prototiles is maintained. A certain kind of point decoration
processes turns out to be useful for the identification of possible
division rules. The method is capable of generating a broad range
of decagonal tilings, many of which are chiral and have atomic
surfaces with fractal boundaries. Two new families of decagonal
tilings are presented; one is quarternary and the other ternary.
Properties of the ternary tilings with rhombic, pentagonal, and
hexagonal prototiles are investigated in detail.
\end{abstract}

\section{Introduction}\label{sec1}

Since the discovery of quasicrystals (QCs) (Shechtman {\it et al.}, 1984),
there has been an up-surge of interest in quasiperiodic tilings with
non-crystallographic point symmetries. Archetypes of such tilings
in the plane are the Ammann-Beenker tiling (Gr\"unbaum \& Shephard, 1987; Beenker, 1982),
the Penrose tilings (Gr\"unbaum \& Shephard, 1987; Penrose, 1974,1978,1979),
and the Stampfli square-triangle tilings (Stampfli, 1986; Baake {\it et al.}, 1992),
which belong to the octagonal, decagonal,
and dodecagonal Bravais classes, respectively. Their vertices form a
quasiperiodic point set called a quasilattice (QL), which mimics the
spatial arrangement of clusters in a two-dimensional (2D) QC. In real
QCs, still a far wider variety of cluster arrangements are possible
under different compositions and temperatures (Edagawa {\it et al.}, 1994; Edagawa {\it et al.}, 2000).
In order to advance our understanding of the real structures, it is
therefore necessary to extend our scope for tiling models of QCs and
to describe the structures in a more systematic fashion. In the present
report, we mean by a 2D tiling {\em a disjoint covering of the plane
by edge sharing copies of a finite number of polygonal prototiles}.

In general, a 2D QL is constructed as a section of a 4D hypercrystal
along the plane (physical space), where the hypercrystal is constructed
from `atomic surfaces' arranged periodically according to the relevant
Bravais hyperlattice; each atomic surface extends only along the
perpendicular space, which is an orthogonal complement to the physical
space. The global characteristics or, more precisely, the
local-isomorphism class of the QL is coded as the size and the shape of
the atomic surface(s) (de Bruijn, 1981; Mackay, 1982; Kramer \& Neri, 1984; Duneau \& Katz, 1985).
\footnote{Local isomorphism classes of QLs are further grouped into
mutual-local-derivability classes (Baake {\it et al.}, 1991; Baake \& Schlottmann, 1997).}

In constructing 2D QLs by the section method, we would practically
be restrained to polygonal atomic surfaces. However, certain important
structures cannot be obtained in this way; the dodecagonal
square-triangle tilings are typical examples which have rather
complicated `fractal' atomic surfaces (Stampfli, 1986; Baake {\it et al.}, 1992; Smith, 1993; Cockayne, 1994).
Further examples with fractal atomic surfaces can also be found in the
literature (Zobetz, 1992; Godr\`eche {\it et al.}, 1993; Niizeki, 2007a).
These structures have been
discovered by inflation methods, in which the shapes of atomic
surfaces are not given {\em a priori}.
It is important to note that even for a known atomic surface with a
fractal boundary, the computation of the structure by the section method
is impractical because minute numerical errors cannot be avoided in 
judging which side of the boundary lies a given point near the boundary.
Therefore, the section method does not suit the generation of such a
QL, whereas an inflation method can work fine.

So far, a systematic attempt has been made for generating 1D binary 
tilings by inflation rules (Luck {\it et al.}, 1993),
revealing that the structures tend to have fractal (Cantor-set like)
atomic surfaces. In more than two dimensions, however, 
inflation rules for quasiperiodic tilings have been rarely known
except for those described above. This situation is simply caused by
a hurdle in finding an inflation rule that does not produce any
inconsistency throughout the entire structure. There is nevertheless
some hope at present; a systematic inflation method for generating
QLs, i.e. quasiperiodic point sets, in general dimensions has been
developed recently (Niizeki, 2008).
Let us call the latter method the
{\em point inflation scheme} (PIS) and each of its inflation
algorithms a {\em point inflation rule} (PIR). At this point, however,
it should be remembered that a QL does not necessarily provide the
vertices of a tiling.

An advantage of employing tiling models for QCs lies in that they
enable the entire structure to be decomposed into a finite number
of {\em prototiles}, or fundamental structural units. Another possible
advantage may lie in their stronger geometrical constraints, which
are based on physical reasonings like the avoidance of unrealistic
short distances. The present aim is therefore to develop a new inflation
scheme for generating quasiperiodic tilings. As in the case of
archetypal tilings in the plane, inflation rules should proceed in
the following steps: 1) an expansion of a tiling and 2) a division of
the expanded tiles into tiles of the original size. Bear in mind that
the expanded tiles can be divided in an arbitrarily way as far as the
set of prototiles is maintained.

In the present report, it is proposed that the division of the tiles
is generalized with the help of a point decoration process.
The point decoration might simply be defined by a PIR, but this cannot
prevent excess points from being generated in general. One needs
to remove a part of the resulting point set so that the remaining points
would be the vertex set of a tiling. Wherefore, the point decoration
can be defined instead by taking an appropriate subset of the set
generated by a PIR. In other words, the PIR is used to obtain the
candidate positions for the decoration, where only a part of it is
adopted. The present scheme turns out to be useful
for discovering unknown tilings. For the sake of a self-contained
presentation, the following arguments are confined to the decagonal
case. Furthermore, the edges of any tiling to be considered are all
given by one of the basic unit vectors,
$\boldsymbol{e}_j$ ($j=0, ..., 9$), which will be defined in Sec.\ref{sec2}.

Sec.\ref{sec2} is devoted to mathematical preliminaries,
in which the decagonal Bravais module $\Lambda_{10}$ is introduced;
it is a projection of the
4D decagonal lattice onto the 2D physical space. The new scheme for
generating decagonal tilings is presented in Sec.\ref{sec3}.
In Sec.\ref{sec4}, the present scheme is exercised and two new
families of decagonal tilings are obtained; one is quarternary and
the other ternary. 
The atomic surfaces of the new tilings are presented in Sec.\ref{sec5}.
In particular, for the ternary tilings, the atomic surfaces with
fractal boundaries are derived geometrically as the
fixed sets of the dual maps associated with the relevant inflation
rules. A statistical analysis is performed for the ternary tilings
in Sec.\ref{sec6}. After presenting further remarks in
Sec.\ref{sec7}, conclusions are given in the final section.

\section{Decagonal Bravais module}\label{sec2}

Let us define the quasicrystallographic axes
for a decagonal QL by the ten unit
vectors $\boldsymbol{e}_j=(\cos{(j\theta)}$, $\sin{(j\theta)})$ with
$j=0, 1, 2, ..., 9$ and $\theta=\pi/5$, pointing at the vertices of a
regular decagon centered on the origin. Among the ten vectors, only
four are linearly independent with respect to integral coefficients.
A conventional set of basis vectors is introduced as
$\boldsymbol{\tilde e}_j:=\boldsymbol{e}_{2j}$ ($j=0,1,2,3$),
pointing at four vertices of a pentagon from the origin, where
the fifth vertex $\boldsymbol{\tilde e}_4:=\boldsymbol{e}_{8}$
is left unused because it can be represented by the four bases;
$\boldsymbol{\tilde e}_4=-\sum_{j=0}^{3}\boldsymbol{\tilde e}_j$.
The basis set generates a Z-module of rank four called the decagonal
Bravais module, denoted by the symbol $\Lambda_{10}$.
Remember that the point group of $\Lambda_{10}$ is the dihedral
group $D_{10}$.
Furthermore, $\Lambda_{10}$ has an important property of scaling
invariance $\tau\Lambda_{10}=\Lambda_{10}$, where $\tau=(1+\sqrt{5})/2$
is the golden mean (Niizeki, 1989a).
\footnote{In general, the ratio $\tau$ is called the Pisot unit in
the algebraic theory of Bravais modules. For the octagonal or the
dodecagonal cases, it takes $1+\sqrt{2}$ or $2+\sqrt{3}$, respectively.}

The decagonal Bravais hyperlattice $\mathcal{L}_{10}$ is defined in a
4D Euclidean hyperspace ${\rm E}_4$, and it is generated by
the primitive basis vectors
$(\boldsymbol{\tilde e}_{j}, \xi\boldsymbol{\tilde e}^{\perp}_{j})$
($j=$ 0,1,2, and 3), where
$\boldsymbol{\tilde e}^{\perp}_{j}:=\boldsymbol{e}_{4j\;({\rm mod}\;10)}$
(Niizeki, 1989a; Yamamoto, 1996)
 and $\xi$ is an arbitrary scale factor satisfying
$0<\xi\ne1$. The point group $G$ of $\mathcal{L}_{10}$ is isomorphic
to the dihedral group $D_{10}$. The 2D physical space ${\rm E}^{\parallel}$
is an irreducible subspace of ${\rm E}_4$ with respect to $G$, and it is
inclined against $\mathcal{L}_{10}$ in an incommensurate fashion.
The orthogonal projection of $\mathcal{L}_{10}$ onto ${\rm E}^{\parallel}$
gives nothing but the decagonal Bravais module $\Lambda_{10}$.

The 2D perpendicular space ${\rm E}^\perp$ is defined as the orthogonal
complement to ${\rm E}^\parallel$ in ${\rm E}_4$; that is,
${\rm E}_4={\rm E}^\parallel \oplus {\rm E}^\perp$. Then the
hyperlattice $\mathcal{L}_{10}$ can be projected onto
${\rm E}^{\perp}$ as well, generating another Z-module,
$\Lambda_{10}^{\perp}$. Since the orthogonal projections from
$\mathcal{L}_{10}$ to both $\Lambda_{10}$ and $\Lambda_{10}^\perp$
are bijections, one can introduce a natural bijection $\hat{\pi}$
between the Z-modules; $\Lambda_{10}^{\perp}=\hat{\pi}\Lambda_{10}$.

We confine ourselves to the case when the vertex set $\Sigma_\mathcal{T}$
of a tiling $\mathcal{T}$ is a subset of the Bravais module $\Lambda_{10}$.
In the ordinary cut-and-projection method (de Bruijn, 1981; Mackay, 1982; Kramer \& Neri, 1984; Duneau \& Katz, 1985),
$\Sigma_\mathcal{T}$ is generated as the orthogonal projection of
$({\rm E}^\parallel+\mathcal{W})\cap\mathcal{L}_{10}$, a cut of
$\mathcal{L}_{10}$ within a strip along E$^{\parallel}$, onto
E$^{\parallel}$. The cross section of the strip,
$\mathcal{W} (\subset {\rm E}^\perp)$, is equivalent to the atomic surface
in the section method.
The image of $\Sigma_\mathcal{T}$ in ${\rm E}^\perp$ is denoted as
$\Sigma_\mathcal{T}^\perp:=\hat{\pi}\Sigma_\mathcal{T}$.
In general, $\Sigma_\mathcal{T}^\perp$ is a dense subset of 
the atomic surface $\mathcal{W}$, which on the other hand should be
included in the closure of $\Sigma_\mathcal{T}^\perp$, i.e. 
$\Sigma_\mathcal{T}^\perp\subset\mathcal{W}\subseteq
\overline{\Sigma_\mathcal{T}^\perp}$.
\footnote{Some authors (Baake {\it et al.}, 1991; Niizeki, 2008)
prefer that an atomic surface $\mathcal{W}$ is a compact
(i.e., closed and bounded) set in E$^\perp$, hence,
$\mathcal{W}=\overline{\Sigma_\mathcal{T}^\perp}$. In this paper,
however, more general language is sought so that a part of the
boundary $\partial\mathcal{W}$ can be absent from $\mathcal{W}$
while fulfilling the glueing condition as explained in the text
for the case of the RPH tilings, see Fig.\ref{fig8}.}

\section{Generalized point processes for tilings}\label{sec3}

In the PIS (Niizeki, 2008),
a PIR is formulated as a set map $\phi$
that acts on an arbitrary subset $L$ of $\Lambda_{10}$.
$\phi$ proceeds by expanding $L$ by a certain ratio $\sigma (>1)$ and
subsequently placing a copy of a certain motif
$\mathcal{S} (\subset \Lambda_{10})$ centered on every vertex.
The ratio $\sigma$ can take any natural power of the Pisot unit $\tau$
(the golden mean) associated with $\Lambda_{10}$, so that $\sigma L$
remains to be a subset of $\Lambda_{10}$. The motif $\mathcal{S}$ is
a bounded set with a finite number of points comprised of one or more
shells, each of which is an orbit of a point with respect to the
point group $D_{10}$. One can put the procedure in a simple form as
$\phi(L):= \sigma L + \mathcal{S}$, in which the $+$ symbol implies
$A+B\equiv\{a+b| ^\forall{a}\in A, ^\forall{b}\in B\}$.
Note that the resultant QL would have the point group $D_{10}$.

The atomic surface of a QL generated by the PIS is identified as the
unique attractor (Hutchinson, 1981)
of the dual set map $\phi^{\perp}$
that acts in the perpendicular space;
$\phi^{\perp}(X)=\sigma^{\ast} X + \mathcal{S}^{\perp}$,
where $\sigma^\ast$ is the algebraic conjugate of $\sigma$ and
$\mathcal{S}^\perp=\hat{\pi}\mathcal{S}$. $\phi^{\perp}$ is nothing but
an iterated function system (IFS), which is a common technique to
generate fractal objects (Falconer, 1990).
Indeed, the atomic surfaces of
many QLs generated by the PIS have fractal boundaries. Various planar
QLs have been found by the PIS (Niizeki, 2008),
which has also been applied to the case of icosahedral QLs (Fujita \& Niizeki, 2008).

Certain QLs can derive a tiling if the points are connected by
uncrossed edges of a unit length, where the tiles are the regions
bounded by these edges.
Let us call this property the {\em unit connectivity} (UC) of the point sets. If a point set
obtained by successively applying a PIR has the UC property, the PIR
can be translated into an inflation rule for the tiling; an expansion of
the tiles followed by their division into the original tiles.
The division rules of the tiles are determined by the generated points
in the tiles. In general, however, undesirable short distances prevent
the generated point sets to be translated into tilings. Such
redundancies can be discarded only by introducing an elimination step at
each iteration to take an appropriate subset which fulfills the UC
property. Then the expanded tiles are divided properly.
We call the new inflation process a generalized point process (GPP).
In the GPP, the role of the PIR is to generate the {\em candidate
positions for the vertices of a tiling}.

For now, the UC property is required by the point set at every iteration
of a GPP. A GPP for decagonal tilings is formally given by the following
steps: Step (I) an expansion of the tiling by the ratio $\sigma$,
where $\sigma$ is a natural power of $\tau$ (the golden mean), and
Step (II) the decoration of every expanded tile by points, where the
positions of the points should originate from $\Lambda_{10}$ and are
assumed to be determined uniquely by the shape of the expanded tile
as well as those of its adjacent neighbours. In practice, candidate positions
for the points of decoration are generated by a PIR, then an appropriate
subset is taken to determine the division of the tiles.
Note that there are degrees of freedom for the choice of the point
decoration provided that it is determined uniquely within the first
adjacent neighbors and that the set of prototiles is maintained.
It is this generality that allows us to produce a broad class of new
quasiperiodic tilings. The degrees of freedom can be promoted further
by allowing the decoration to depend on farther neighbors. However,
this would complicate the algorithm and will not be considered.

\section{Examples of decagonal tilings}\label{sec4}

Several decagonal tilings are generated by the GPP scheme. All the tilings
presented below obey the UC property, in which the vertices are connected
through the ten basic unit vectors $\boldsymbol{e}_j$ ($j=0-9$) forming
the edges of the tiles. Any vertex can therefore be written as
$\boldsymbol{l}=n_0\boldsymbol{\tilde e}_0+n_1\boldsymbol{\tilde e}_1+
n_2\boldsymbol{\tilde e}_2+n_3\boldsymbol{\tilde e}_3$, or alternatively
it can be indexed as $\boldsymbol{l}=[n_0 n_1 n_2 n_3]$.

\subsection{Para-Penrose tiling}\label{subsecPP}

The present tiling has been already reported (Niizeki, 2008)
as one of the simplest decagonal tilings that can be generated by the PIS.
The scaling ratio of the relevant PIR is $\sigma=\tau^2$, while the
motif $\mathcal{S}$ consists of two shells $\langle[0000]\rangle$ and
$\langle[1100]\rangle$,
where $\langle\boldsymbol{l}\rangle:=\{g \boldsymbol{l} | g\in D_{10}\}$
is the orbit of
the point $\boldsymbol{l}$ with respect to $D_{10}$. In addition to the
four prototiles,
i.e., the 36$^\circ$ rhombus (R), the regular pentagon (P), the crown (C),
and the pentacle star (S), of the pentagonal Penrose tiling (P1)
(Penrose, 1974; Gr\"unbaum \& Shephard, 1987),
it has an additional prototile, namely the barrel shaped
hexagon (H). This tiling is called the {\em para-Penrose tiling} by
Niizeki (2008).

\begin{figure}
\includegraphics[width=8cm]{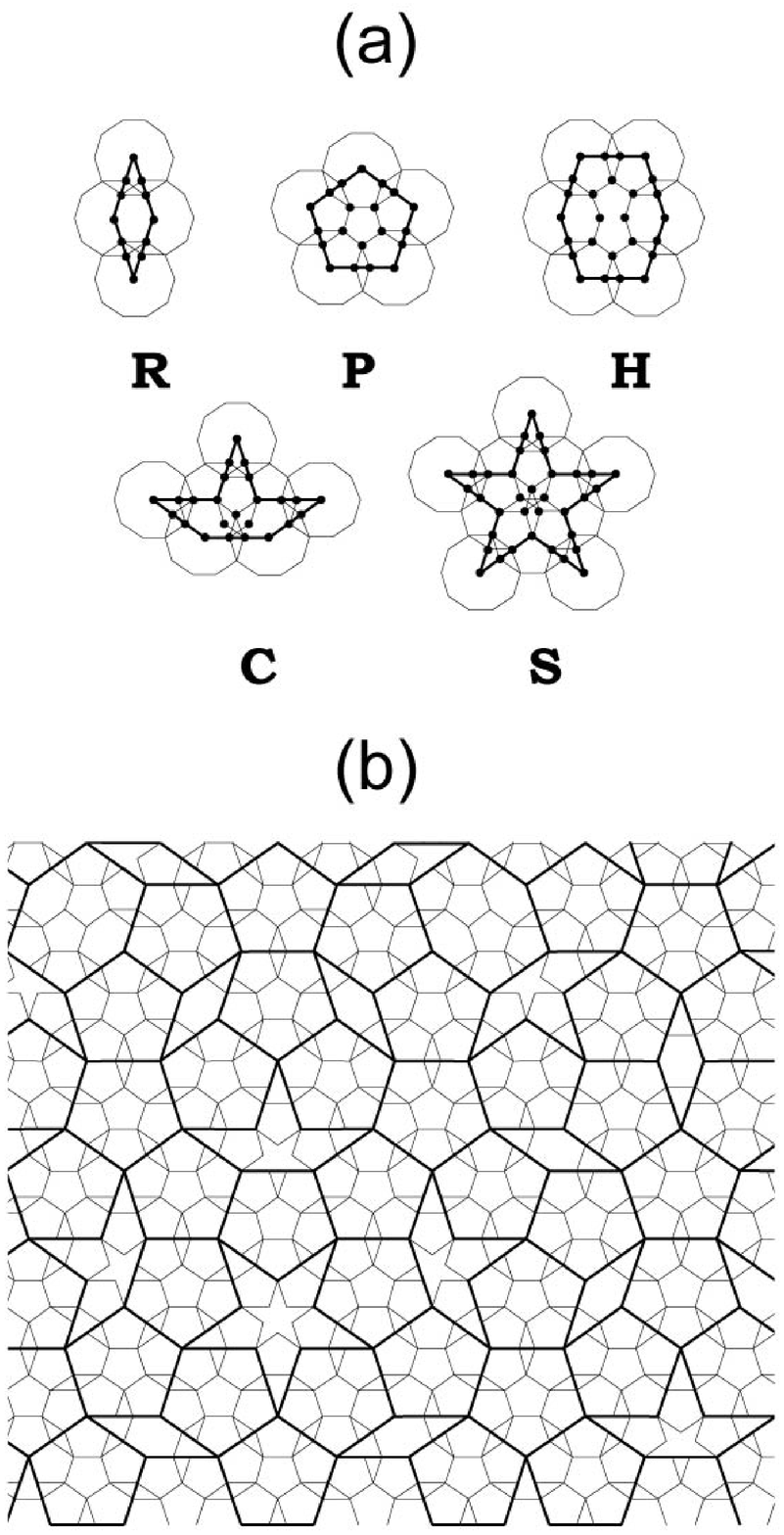}
\caption{(a) The point decorations of the expanded prototiles
for the para-Penrose tiling are determined by superposing
decagons centered at all the vertices.
(b) A square patch of the para-Penrose tiling with expanded
tiles being indicated with the thicker lines shows the
division rules for the expanded tiles.}
\label{fig1}
\end{figure}

In a single iteration, these tiles are expanded by the ratio $\tau^2$
and then a copy of $\mathcal{S}$ is placed on every vertex;
see Fig.\ref{fig1}(a). One finds that each expanded tile
is uniquely decorated with the points of $\mathcal{S}$'s within its
border. As the new points are connected with unit edges, the
expanded tile is divided into tiles of the original size, whereas
in the vicinity of the boundary the tiles can be shared with
the neighboring expanded tiles. In particular, when dividing an
expanded P- or H-tile, segments near the boundary can be attributed
to different kinds of tiles. This means that the division rule
is not unique within the expanded tile itself. Still, it turns out
to be unique within the first adjacent neighbors as one can readily
check in Fig.\ref{fig1}(b). Therefore, the present PIR satisfies
the requisites for a GPP.

\subsection{RPHC tilings}

The PIR for the para-Penrose tiling can be slightly altered by removing
a generated vertex inside every expanded C-tile at the symmetrical
position. Obviously this does not affect the uniqueness of the
decoration within every expanded prototile, and the resulting
point set is shown to satisfy the UC property. As a consequence,
the combination of an
S-tile and a P-tile (S-P complex) lying in the center of every expanded
C-tile turns into one of a C-tile and an H-tile (C-H complex).
Since the S-prototile may no longer appear except in the center of an
expanded S-tile, it is a marginal one in the present case; that is, an
iteration of the above GPP will generate a quaternary tiling with R-,
P-, H-, and C-prototiles as shown in Fig.\ref{fig2}(a). The division
rule of every expanded tile depends on its first adjacent neighbors as
in the case of the para-Penrose tiling.

There are yet different ways to remove a vertex in an expanded C-tile
without affecting the set of prototiles. Let us consider a
gear-shaped complex (or a G-complex),
including an S-tile in the center and five adjacent P-tiles
(see Fig.\ref{fig3}(a)), lying on the bottom of the expanded C-tile.
One or two of its five internal vertices can be removed so that the
G-complex is to be divided in different manners. By removing
one vertex, the G-complex is divided into one H-, one C-, and four
P-tiles (Fig.\ref{fig3}(b), left). There are five possible
orientations, one of which has been taken in the preceding paragraph.
Two other orientations are used in Fig.\ref{fig2}(b) and (c),
while the remaining two are their mirror images.
Removing two internal vertices offers yet different ways to divide
the G-complex. Of the five possible orientations, the only
one without breaking the mirror symmetry is used in Fig.\ref{fig2}(d).
Two other choices are shown in Fig.\ref{fig2}(e) and (f), while the
remaining two are their mirror images.
By removing two vertices, the G-complex is divided into one R-,
two H-, and three P-tiles (Fig.\ref{fig3}(b), right).

We have constructed in total ten GPPs for tilings with the identical
set of four prototiles. Recall that the eight GPPs without mirror
symmetry would generate chiral tilings. If the ten GPPs
are applied in a mixed and arbitrary order, infinitely many tilings
can be generated. Furthermore, if different GPPs are allowed to be
applied at different locations at the same time, the possibilities
become illimitable.

\subsection{RPH tilings}\label{subsecRPH}

Let us argue how the C-prototile can be excluded from the prototiles of 
RPHC tilings. Consider a turban-shaped complex (or a T-complex)
composed of a C-tile and two adjacent P-tiles in an RPHC tiling.
A T-complex is associated to every 36$^\circ$ angle
of the expanded R- and C-tiles. The T-complex can be decomposed into
three tiles, R-, P-, and H-tiles, by removing one of the two inner
vertices, thereby breaking the mirror symmetry; see Fig.\ref{fig3}(c).
If all the T-complexes
are (re-)divided in such a way, an RPHC-tiling turns into a tiling
which does not include any C-tile. The two different ways to divide
the T-complex are mirror images of one another, where {\em left}-
and {\em right}-handed chiralities are associated with them.

It is tempting to apply either one of the chiral rules for dividing all
the T-complexes; then the two GPPs for both chiralities are obtained.
These GPPs maintain the set of the prototiles, R, P, and H.
The left-handed GPP is shown in
Fig.\ref{fig4}, the repetition of which generates a tiling shown in
Fig.\ref{fig5}(a). The mirror images of these figures corresponds to
the right-handed counterparts.
Note that the point decorations within the expanded prototiles are not
uniquely determined in this case, but they are unique within the
first adjacent shell. The decoration uniquely determines the division of
the expanded prototiles. The two GPPs may be applied in an
arbitrary order, resulting in an infinite number of RPH tilings.
A tiling shown in Fig.\ref{fig5}(b) is an outcome of an alternate
repetition of the right- and the left-handed GPPs, where the final
GPP is the left-handed one. In both of the tilings shown in
Fig.\ref{fig5}, a spiral pattern associated with the left-handed chirality
can be easily recognized.
Again, much greater possibilities exist if the two GPPs are allowed to
be applied at different locations at the same time.

\section{Atomic surfaces}\label{sec5}

The atomic surface of a quasiperiodic tiling $\mathcal{T}$ is closely
connected with the global characteristics of the structure.
It can be inferred by projecting the vertices of a patch of
$\mathcal{T}$ containing sufficient number of vertices onto
the perpendicular space, E$^\perp$. This is done in Fig.\ref{fig6}
for the six RPHC tilings and in Fig.\ref{fig7} for the two RPH tilings,
respectively. The convex hulls of all these atomic surfaces are the 
regular decagon whose vertices are given by $\xi\boldsymbol{e}_{j}$
($j=0,...,9$), corresponding to the atomic surface of the Penrose P1
tiling. However, different types of erosion are observed near the
boundaries even for tilings with the same set of prototiles.
The symmetry of an atomic surface reflects that of the relevant tiling;
all the atomic surfaces except the two with the mirror symmetric GPPs
(Fig.\ref{fig6}(a) and (d)) have the lower point symmetry with the
cyclic group $C_{10}$. For each case, the erosion reveals a complicated
fractal patterns. In particular, the three atomic surfaces
that are shown in Fig.\ref{fig6}(a),(c), and (e) exhibit hierarchical
holes in the boundary regions. The rest of the atomic surfaces maintain
the disc-like topology.

Recall that, for a QL generated by the PIS, the atomic surface
is simply identified as the unique attractor of the dual set map
$\phi^\perp(X)=\sigma^\ast X + \mathcal{S}^{\perp}$ in E$^\perp$, where
$\sigma^\ast$ denotes the algebraic conjugate of $\sigma$ and $X$ any
subset of E$^\perp$ (Niizeki, 2008).
For a tiling generated by the GPP
scheme, on the other hand, the dual set map is associated with the
determination of the candidate positions, from which a certain subset
should be eliminated. The elimination is represented in E$^\perp$ as
a subtraction of unnecessary parts from $\phi^\perp(X)$.
In the following, the latter process is presented for the RPH tilings,
in which case geometrical rules to determine the atomic surfaces
are identified. However, since such geometrical rules are not easy
to generalize, the case of the RPHC tilings will be left open.

The PIR for the para-Penrose tiling given in Subsec.\ref{subsecPP}
gives $\mathcal{S}^\perp=\{\langle[0000]\rangle,\langle[1010]\rangle\}$
and $\sigma^\ast=\tau^{-2}$ for the dual set map $\phi^\perp$.
The fixed set of $\phi^\perp$, which is a moth-eaten version of
the regular decagon (Niizeki, 2008),
is the corresponding atomic
surface. The small part that is left out from the decagonal
atomic surface corresponds to a portion of concave vertices
of C-tiles and S-tiles, resulting in the emergence of H-tiles
in the para-Penrose tiling.

In order to obtain the atomic surfaces for the RPH tilings, a
subtraction process needs to be combined with the dual set map at
each iteration. In Fig.\ref{fig8}, the initial polygon $X_0$ is defined
as a decagonal star, which is known to be the atomic surface of a
non-chiral RPH tiling (Papadopolos \& Kasner, 2003).
It is transformed by the
dual set map to $\tilde{X_0}=\phi^\perp(X_0)$.
The subtraction process for the chiral GPPs (Subsec.\ref{subsecRPH})
can be understood as a carving process near the boundary of
$\tilde{X_0}$. Note that five strips are superposed on the next
figure, while five more strips can be superposed up-side-down but
are suppressed. These strips are cut
by the boundary of $\tilde{X_0}$, while the two ends of each strip
do not coincide through a translation. Accordingly, within each cut
of the strips,
two points can lie on a line parallel to the strip with the distance
$\xi\tau$. This leads to an excessive appearance of short distance
$1/\tau$ in E$^\parallel$ and hence to the existence of C-tiles.
It turns out that for an RPH-tiling no such pair of
points is allowed to exist within a single strip. Therefore,
either or both ends of each cut must be carved so that they 
coincide through $\xi\tau$-translations.

The carving process for the left- or right-handed GPP is simply
to carve a single end only. As shown in Fig.\ref{fig8}, five strips
are arranged to form a pentacle star, which can be traced either
clock-wise or anti-clock-wise. Let us fix that the pentacle should
be traced clock-wise. Then for the left-handed GPP, the first end
of each strip encountered while tracing the pentacle should be
carved so that it coincides with the translate of the second end.
These two ends fulfill the gluing condition; namely, if a point on
one end is taken into account, the corresponding point on the other
end should be discarded.
The carving process for all the ten strips results in a new polygon
$X_1$, which is the atomic surface of another RPH-tiling (chiral in
this case). The carving process is denoted as $\gamma_l$, while
the right-handed counterpart is denoted as $\gamma_r$ in which case
the opposite end of each strip is to be carved.

The atomic surface of an RPH-tiling generated by the GPP scheme can
be obtained by repeatedly applying $\gamma_l\cdot\phi^\perp$ and/or
$\gamma_r\cdot\phi^\perp$. In the example
shown in Fig.\ref{fig8}, only the left-handed process 
$\gamma_l\cdot\phi^\perp$ is used.
In general, the subtraction process $\gamma$ associated with the
removal of unnecessary points from the candidate positions can be
rather complicated. At present, the author has been able to identify
$\gamma$ only for the chiral RPH-tilings.

\section{Statistics of the RPH tilings}\label{sec6}

The inflation matrices are introduced in the following to analyze
the statistics of the RPH tilings. For each chirality, there are
six types of tiles that are divided differently, as can be seen in
Fig.\ref{fig4}. In order to define the inflation matrix, the
generated tiles need to be classified further into six types, so
that their numbers can be counted in each expanded tile. Note,
however, that the latter task depends on which chirality is to be
used in the next iteration. In Fig.\ref{fig4}, some of the small tiles
are assigned different letters according to the chirality of the
next GPP. There are four pairs of chiralities for the two
successive iterations, $ll$, $rl$, $lr$, and $rr$.
It is sufficient to consider two particular cases, $ll$ and $rl$,
since the other cases are mere mirror images of the former two;
the latters have the same inflation matrices as the formers.

In the case $ll$, the division rules shown in Fig.\ref{fig4} reveals
that the six expanded tiles are divided as
\begin{eqnarray}
&&\begin{array}{l}
V_A=V_a+\frac{1}{\tau^2}V_d+V_e,\\
V_B=\frac{5}{2}V_a+V_b+5V_c,\\
V_C=2V_a+2V_b+3V_c+\left(1-\frac{1}{2\tau^2}\right)V_d,\\
V_D=\frac{5}{2}V_a+3V_b+2V_c+\left(3-\frac{1}{2\tau^2}\right)V_d,\\
V_E=2V_a+4V_b+\left(4-\frac{1}{\tau^2}\right)V_d,\\
V_F=3V_a+2V_b+4V_c+2V_d,
\end{array}
\end{eqnarray}
where $V_x$ represents the volume of the tile labeled `$x$'.
In the case $lr$, the corresponding formulae are
\begin{eqnarray}
&&\begin{array}{l}
V_A=V_{\bar{a}}+\frac{1}{\tau^2}V_{\bar{d}}+V_f,\\
V_B=\frac{5}{2}V_{\bar{a}}+V_b+5V_{\bar{c}},\\
V_C=2V_{\bar{a}}+V_b+4V_{\bar{c}}+\left(1-\frac{1}{2\tau^2}\right)V_{\bar{d}},\\
V_D=\frac{5}{2}V_{\bar{a}}+2V_b+3V_{\bar{c}}+
\left(3-\frac{1}{2\tau^2}\right)V_{\bar{d}},\\
V_E=2V_{\bar{a}}+2V_b+2V_{\bar{c}}+\left(4-\frac{1}{\tau^2}\right)V_{\bar{d}},\\
V_F=3V_{\bar{a}}+2V_b+4V_{\bar{c}}+2V_{\bar{d}}.
\end{array}
\end{eqnarray}
It follows that the inflation matrices for all the four cases are given by
\begin{eqnarray}
\boldsymbol{M}_1 &:=& \boldsymbol{M}_{ll} = \boldsymbol{M}_{rr} \nonumber\\
&=& \left(
\begin{array}{cccccc}
1 & 0 & 0 & \frac{1}{\tau^2} & 1 & 0\\
\frac{5}{2} & 1 & 5 & 0 & 0 & 0\\
2 & 2 & 3 & 1-\frac{1}{2\tau^2} & 0 & 0\\
\frac{5}{2} & 3 & 2 & 3-\frac{1}{2\tau^2} & 0 & 0\\
2 & 4 & 0 & 4-\frac{1}{\tau^2} & 0 & 0\\
3 & 2 & 4 & 2 & 0 & 0
\end{array}
\right), 
\end{eqnarray}
and
\begin{eqnarray}
\boldsymbol{M}_2 &:=& \boldsymbol{M}_{lr} = \boldsymbol{M}_{rl} \nonumber\\
&=& \left(
\begin{array}{cccccc}
1 & 0 & 0 & \frac{1}{\tau^2} & 0 & 1\\
\frac{5}{2} & 1 & 5 & 0 & 0 & 0\\
2 & 1 & 4 & 1-\frac{1}{2\tau^2} & 0 & 0\\
\frac{5}{2} & 2 & 3 & 3-\frac{1}{2\tau^2} & 0 & 0\\
2 & 2 & 2 & 4-\frac{1}{\tau^2} & 0 & 0\\
3 & 2 & 4 & 2 & 0 & 0
\end{array}
\right). 
\end{eqnarray}
One can check that the maximal eigenvalues of both $\boldsymbol{M}_1$ and
$\boldsymbol{M}_2$ are $\tau^4$, corresponding to the rate of volume increase
under a single GPP iteration. The relevant right-eigenvectors are
common to the two matrices, reading
\begin{eqnarray}
\boldsymbol{v}&=&(\tau^3-4,\frac{1}{2}\left(1+\frac{1}{\tau^2}\right),
\frac{1}{2}\left(1+\frac{1}{\tau^2}\right),1,1,1)^t \nonumber\\
&\doteqdot&(0.236, 0.691, 0.691, 1,1,1)^t,
\end{eqnarray}
where the superscript $t$ indicates the transposition. These six
components provide the relative area (volume) of the tiles labeled
`$a$' to `$f$'.

On the other hand, the left-eigenvectors for the common maximal
eigenvalue $\tau^4$ give the number ratio of the tiles. For $\boldsymbol{M}_1$
and $\boldsymbol{M}_2$, the relevant left-eigenvectors are
\begin{eqnarray}
\boldsymbol{u}_1&=&(\tau^4, \tau^3+1, 2\tau^3, \tau^3-1, 1, 0) \nonumber\\
&\doteqdot&
(6.854, 5.236, 8.472, 3,236, 1, 0),\\
\boldsymbol{u}_2&=&(\tau^4, \tau^3-1, 2\tau^3+2, \tau^3-1, 0, 1) \nonumber\\
&\doteqdot&
(6.854, 3.236, 10.472, 3,236, 0, 1),
\end{eqnarray}
respectively.
That the eigenvectors are different might seem
troublesome if the two GPPs are applied in an arbitrary order.
It turns out, however, that no problem is caused by the difference,
since the inflation matrices multiplied by the wrong eigenvectors
will just give the right ones, i.e.,
\begin{eqnarray}
\boldsymbol{u}_1\boldsymbol{M}_2 = \tau^4 \boldsymbol{u}_2,\\
\boldsymbol{u}_2\boldsymbol{M}_1 = \tau^4 \boldsymbol{u}_1.
\end{eqnarray}
It follows that the number ratios of the tiles depend only on the
chiralities of the final two iterations.
The above two distributions for the six types of tiles do not
cause a difference in the number ratios of the three prototiles,
since the sum over components for P-tile or H-tile are common to
the two left-eigenvectors. It follows that the number ratios of
the three prototiles in the present RPH-tilings are
$R:P:H=1:2:1/\tau$, while the mean volume of the tiles is $1/\tau$
times that of an H-tile.

The differences in the left-eigenvectors manifest themselves in
the statistics of local arrangements of tiles in the relevant
tilings. For instance, the second components of $\boldsymbol{u}_1$ and
$\boldsymbol{u}_2$ representing the relative frequencies of P-tiles
labeled `$b$' are different. This is manifested in the frequencies
of the local centers of five-fold symmetry in the two tilings
shown in Fig.\ref{fig5}; the difference can be rather significant
from the viewpoint of the structural stability as well as the
physical properties if these tilings are to be used for modeling
physical QCs. Importantly, the difference should also be connected
to the boundary shapes of the atomic surfaces, which point will
be left for a future investigation though.
\footnote{Identifying the classes of local centers of symmetry (Niizeki, 1989b)
in quasiperiodic tilings is an important task, which has been
addressed in a systematic way for the cases with polygonal atomic
surfaces (Niizeki, 2007b).
A generalized argument is hence necessary to handle the cases with fractal atomic surfaces.}
There are on the other hand three types of local centers of
two-fold symmetry (Niizeki, 1989b)
located (i) at the centers of R-tiles labeled `$a$' or `$\bar{a}$',
(ii) at the centers of H-tiles labeled `$e$', `$\bar{e}$', or `$f$',
and (iii) at the mid-edge positions between two adjacent P-tiles
both labeled `$b$'. The frequencies of these two-fold centers
are common to the two cases.

\section{Further remarks}\label{sec7}

The generality of the GPP scheme is capable of generating a number
of unknown quasiperiodic tilings, many of which has a chirality.
Let us briefly consider how the structure factor of a chiral 
tiling looks like. In our examples, the breaking of the mirror
symmetry is carried by a small part of the atomic surface near
the boundary, while the main body of the atomic surface maintains
the mirror symmetry. Since the latter part provides the main
contribution to the structure factor, the chirality is only
manifested in relatively weak Bragg peaks. A structure factor is
shown in Fig.\ref{fig9} for the tiling shown in Fig.\ref{fig5}(a),
assuming a point scatterer on every vertex.

In the GPP scheme, the removal of unnecessary points from the
given candidates for the point decoration forms a critical
step. This makes the GPP the most general and robust technique
for generating decagonal tilings. Furthermore, the application
of the basic idea to the octagonal as well as the dodecagonal
cases is straightforward. Indeed, for most of the known tilings,
whether the atomic surfaces are polygonal or fractal, inflation
rules can be re-phrased as GPPs, i.e. the combination of an
expansion step and a point decoration step.

Let us consider the particular case of the dodecagonal 
square-triangle tilings. In this case, each of the vertices
in the expanded tiling is decorated by a three shell motif
$\mathcal{S}$, comprising the origin, an inner hexagon,
and an outer dodecagon. The hexagon can take two different
orientations, which can be chosen at random (Smith, 1993; G\"ahler, 1988)
or according to a deterministic rule; for instance, the local
coordination can be used to fix the orientation (Hermisson {\it et al.}, 1997).
One can see a similarity of the situation to the case of
decagonal tilings that are presented in this paper.

Different tilings that are generated by applying different 
GPPs in different orders are all members of the random tiling
ensemble with the same set of prototiles. They are likely to
have energies very close to each other, so do they have
similar statistical weights. Since they form a deterministic
subset of the relevant random tiling ensemble, the relevant
contribution to the entropy is called the `deterministic
entropy', a term coined by A. P. Smith (Smith, 1993).

Either of the left-handed and the right-handed GPPs for the
RPH tilings is defined by exclusively applying one of the two
division rules for the T-complex; see Fig.\ref{fig3}(c).
One readily recognizes that these two division
rules are mutually connected through a phason flip involving
three tiles. Accordingly, the RPH tilings embrace an abundance
of flipping sites. This has a significant implication when
a physical realization of a similar structure is to be considered,
since the phason degrees of freedom must play an important
role in the structural stabilization. One should also bear
in mind that the same kind of phason flips has been observed
{\em in situ} in a $d$-Al-Cu-Co QC at 1123 K with a
transmission electron microscope (Edagawa {\it et al.}, 2000).

\section{Conclusions}\label{sec8}

A general inflation scheme for generating decagonal quasiperiodic
tilings has been proposed.
In the new scheme, inflation rules are practically comprised of three
steps: an expansion with the ratio $\sigma$, a decoration of every
vertex by a finite motif $\mathcal{S}$, and an elimination of unnecessary
points by local rules. At every iteration, the resulting point set
should be unit-connective and form the vertex set of a tiling. Since
the point decoration process can be readily generalized, various new
tilings are expected to be found. The usefulness of the present scheme
has been demonstrated by generating several new decagonal tilings,
among which the family of ternary tilings has been analyzed in detail.
The concept can be exported not only to the octagonal and dodecagonal
cases, but also to the icosahedral cases (including P-, F-, and I-type
Bravais classes), for which only a few tilings have been known.

\ack{The author is greatly indebted to K. Niizeki for instructions in the
mathematics of quasilattices and for a comment on the issue of
local centers of symmetries in the present tilings.
He is also grateful to T. Ogawa for a comment on spiral patterns in
the chiral RPH tilings.}


\newpage

\newpage

\begin{figure}
\includegraphics[width=7.5cm]{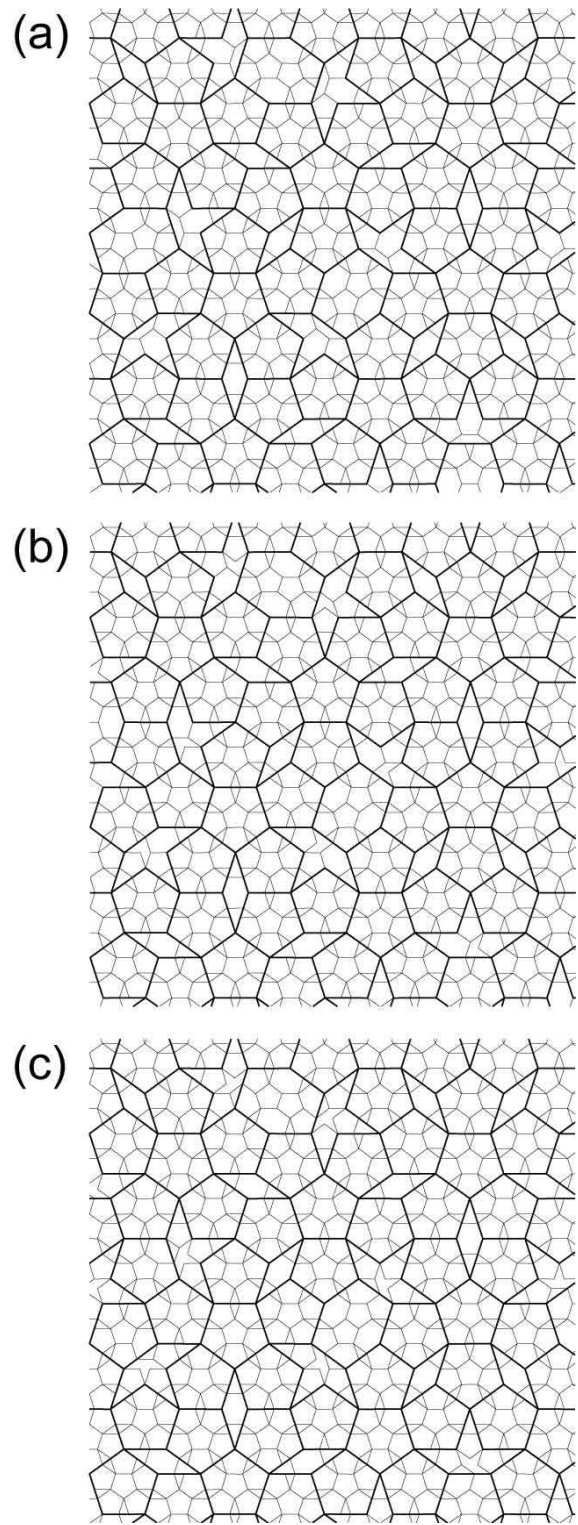}
\caption{Square patches of RPHC tilings generated by GGPs 
with expanded tiles being indicated with the thicker lines.
They are distinguished by the division of the G-complex
lying on the bottom of every expanded C-tile. The division
rules of the expanded P- and H-tiles are also affected by
adjacent C-tiles. (a-c) A single point inside every G-complex
is eliminated, while the mirror simmetry is retained only in (a).}
\label{fig2}
\end{figure}

\addtocounter{figure}{-1}
\begin{figure}
\includegraphics[width=7.5cm]{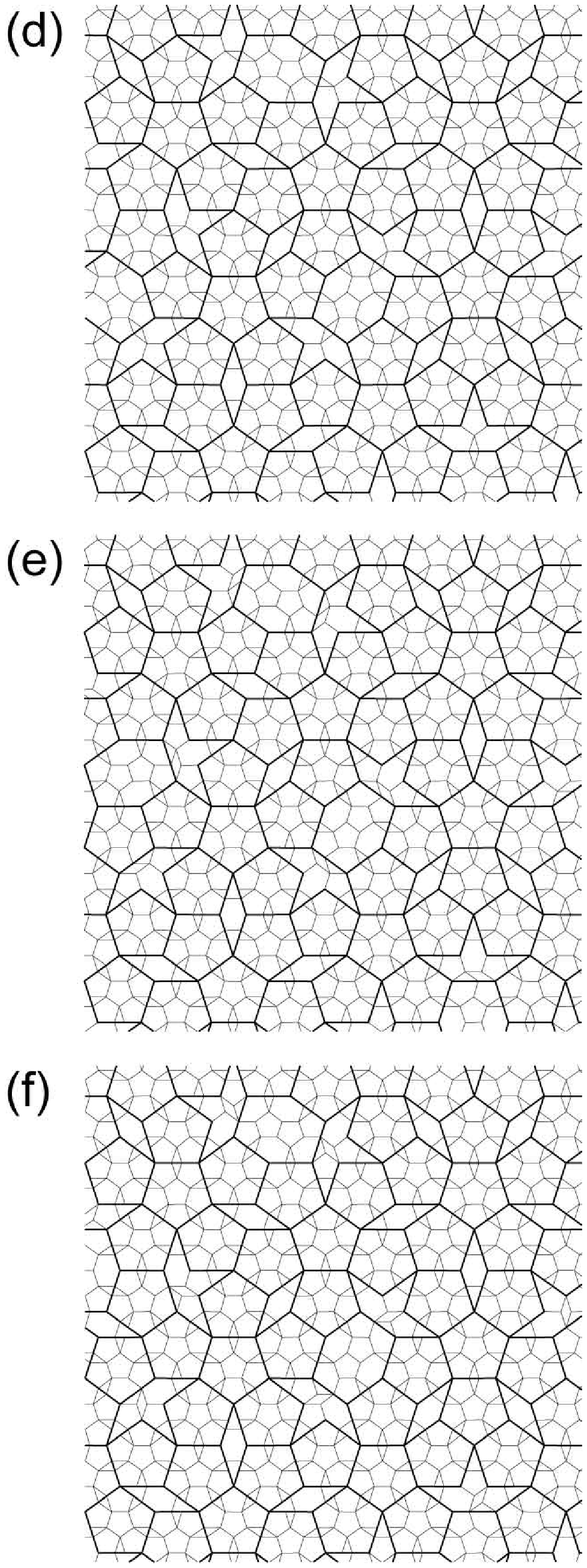}
\caption{(Continued.) (d-e) Two internal points of every G-complex
are eliminated, while the mirror symmetry is retained only in (d).}
\end{figure}

\newpage

\begin{figure}
\includegraphics[width=6cm]{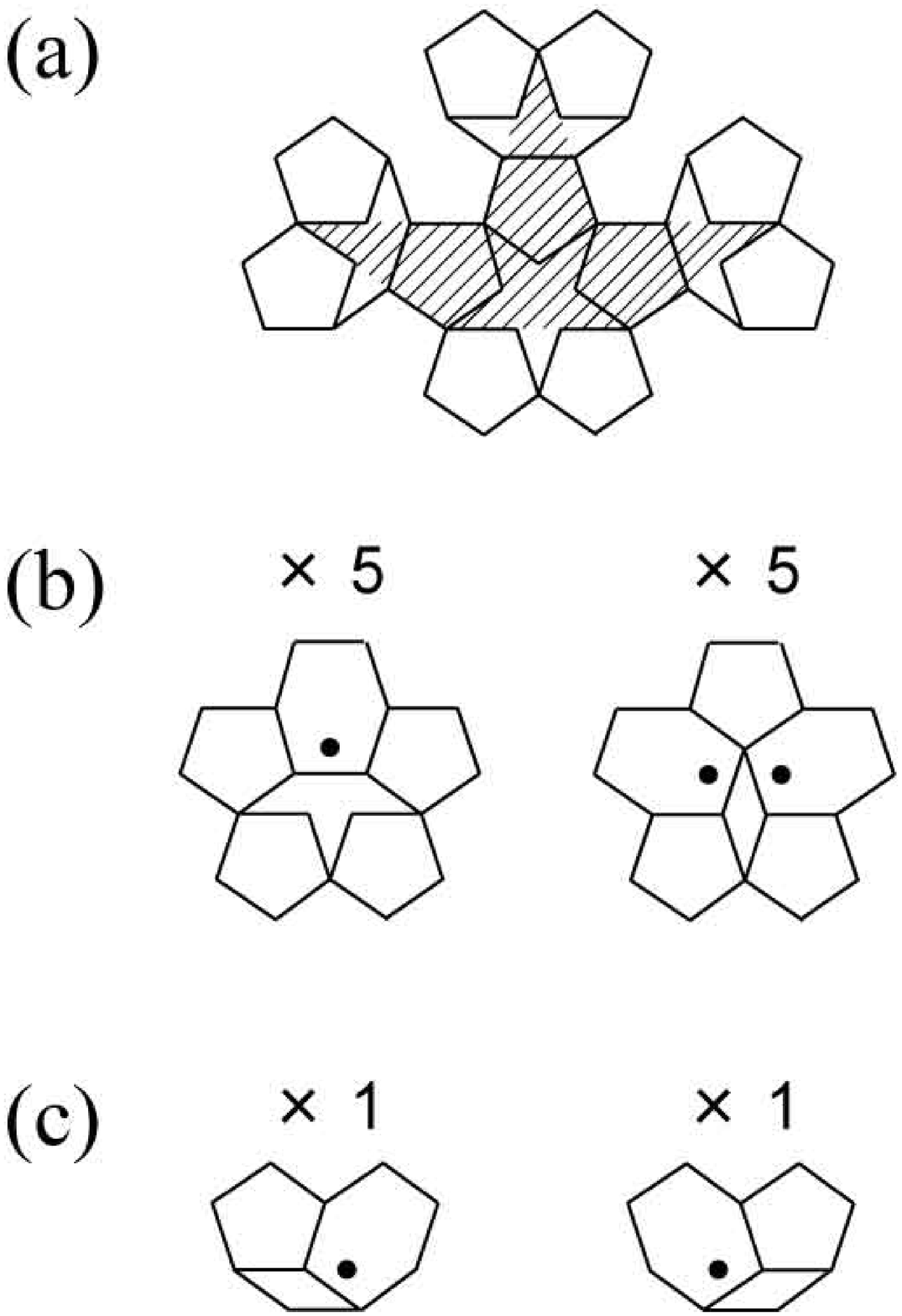}
\caption{(a) One G- and three T-complexes associated with an
expanded C-tile (the hatched area) are shown.
(b) Depending on whether one or two internal vertices
are removed,
a G-complex is divided in two different ways, which may further
take five distinct orientations relatively to the expanded C-tile.
(c) A T-complex can be divided in two ways depending on the
choice of an internal vertex to be removed. In (b) and (c),
the black dots indicate the removed vertices.}
\label{fig3}
\end{figure}

\newpage

\begin{figure}
\includegraphics[width=8cm]{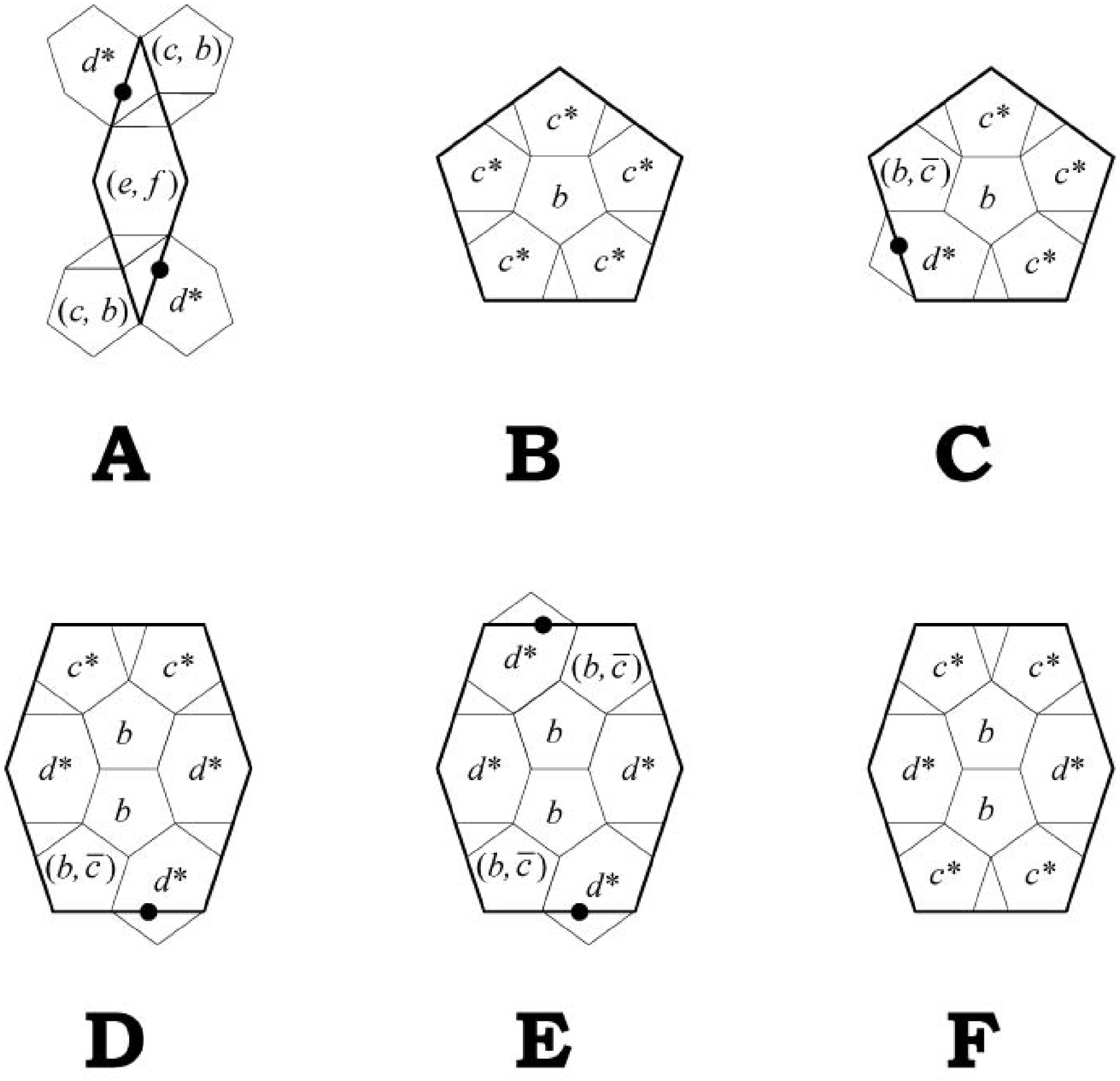}
\caption{The left-handed GPP for the RPH tilings is shown.
An expanded P-
or H-tile is divided in several different ways depending on its first
adjacent neighbors. Expanded tiles are thus classified into six
classes according to their division rules, and they are labeled
`A' to `F'. The black dots indicate the removed vertices inside
T-complexes. The generated tiles can also be classified into
six classes but in two different ways depending on the chirality of
the subsequent GPP (see also Sec.\ref{sec6}). For instance, a P-tile
labeled `($b$, $\bar{c}$)' implies that it is
classified into the $b$-subclass if the next GPP is left-handed while
into the $\bar{c}$-subclass if the next GPP is right-handed, where
the bar indicates the mirror inversion. A mutually inverted pair is
denoted by a starred label; that is, $c^\ast=(c,\bar{c})$ and
$d^\ast=(d,\bar{d})$. The label `$a^\ast$' (i.e. `$(a,\bar{a})$') for
each R-tile is suppressed merely for the reason of space.
Tiles with the label `$f$' only appear if the two successive GPPs
have opposite chiralities, while tiles with the label `$e$'
(or `$\bar{e}$') only appear if the same GPP is applied successively.}
\label{fig4}
\end{figure}

\newpage

\begin{figure}
\includegraphics[width=7.5cm]{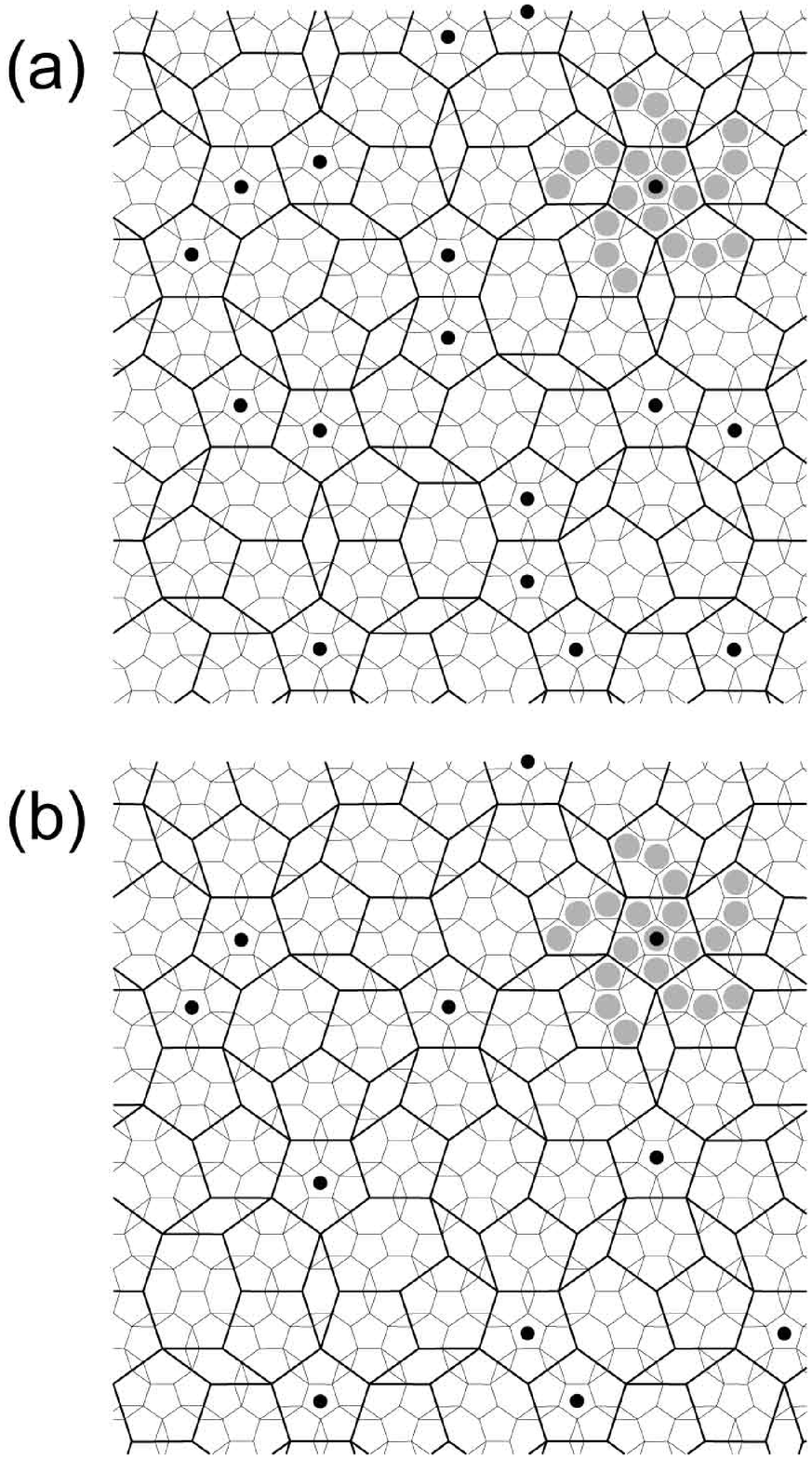}
\caption{Square patches of RPH tilings generated by repeating the
left-handed GPP (a) and by alternating between the right- and
left-handed GPPs (b). Spirals in the arrangement of P-tiles are
emphasized with the gray spots. The black dots indicate the centers
of the expanded P-tiles with the five-fold symmetric division rule
(Fig.\ref{fig4}-B), showing that the frequencies of local five-fold
centers are somewhat different between the two.}
\label{fig5}
\end{figure}

\newpage

\begin{figure}
\includegraphics[width=7.5cm]{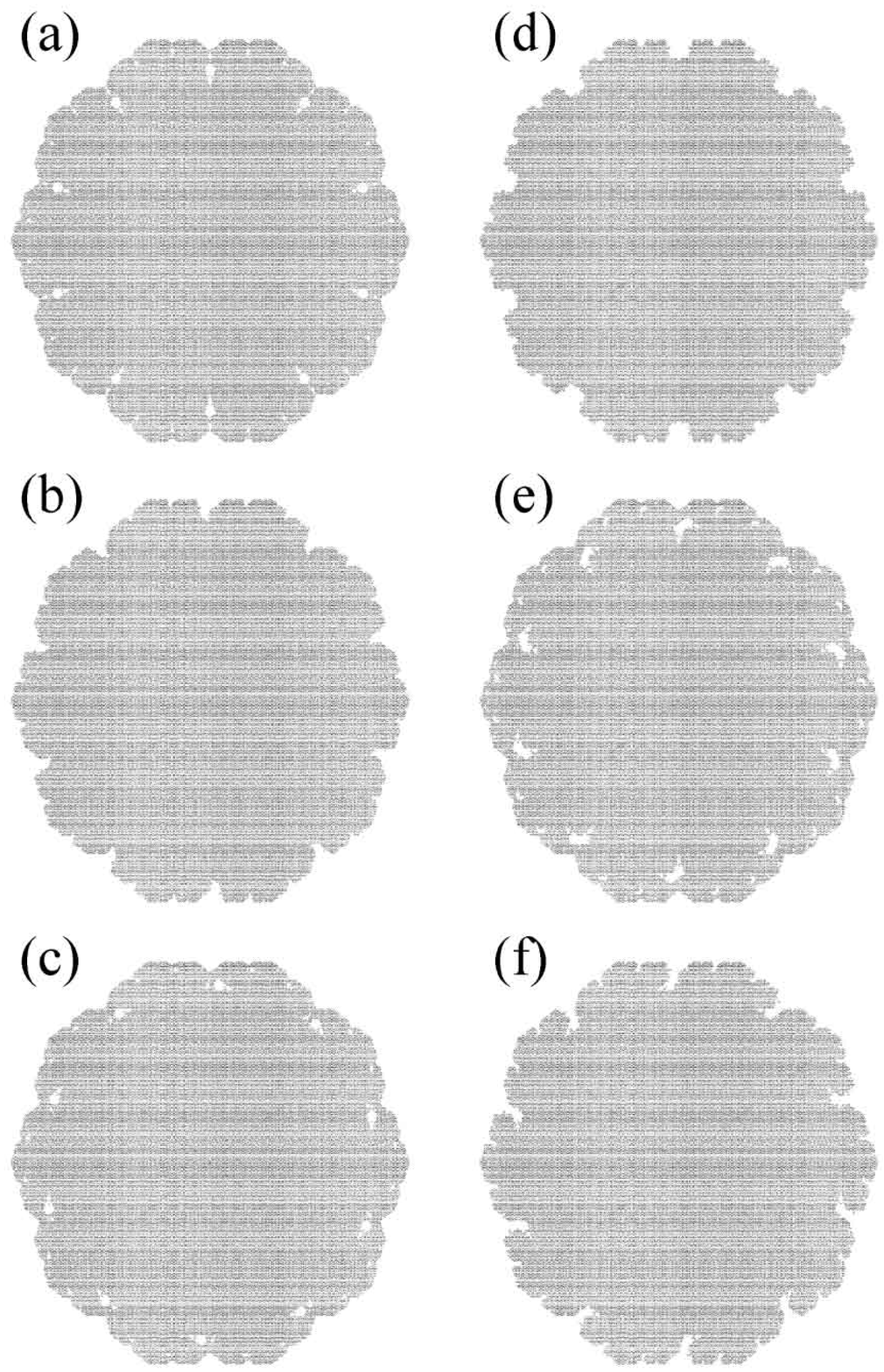}
\caption{(a)$\sim$(f): The atomic surfaces for the RPHC tilings in
Fig.\ref{fig2}(a)$\sim$(f), respectively. The gray areas represent
the projections of a large patch containing over 250,000 vertices
onto E$^{\perp}$. Fine details including hierarchical holes (pits)
are visible.}
\label{fig6}
\end{figure}

\newpage

\begin{figure}
\includegraphics[width=5cm]{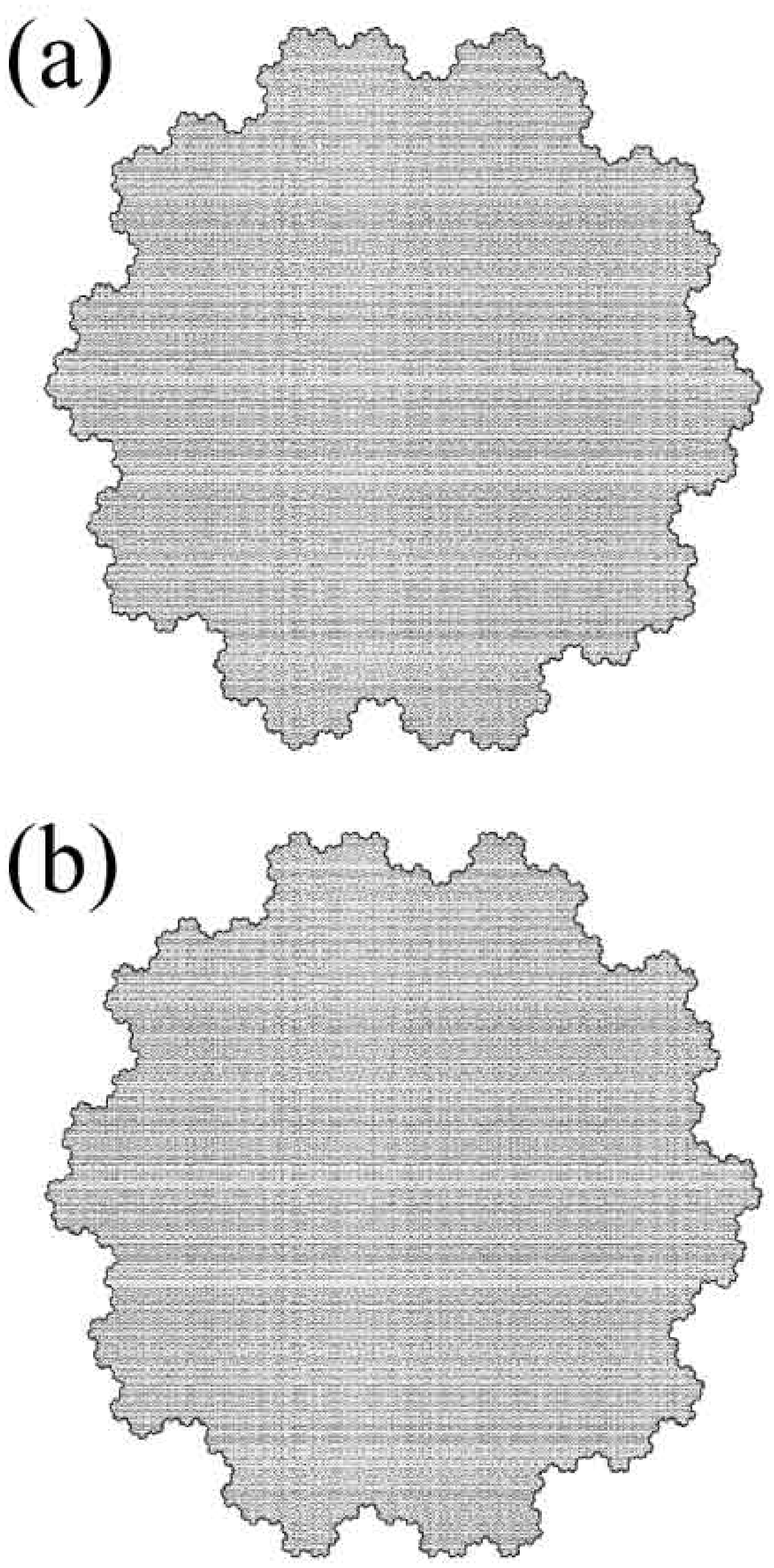}
\caption{(a) and (b): The atomic surfaces for the RPH tilings in
Fig.\ref{fig5}(a) and (b), respectively. The gray areas represent
the projections of a large patch containing over 240,000 vertices
onto E$^{\perp}$, while the boundary lines are obtained with the
geometrical rules as presented in Fig.\ref{fig8}.}
\label{fig7}
\end{figure}

\newpage

\begin{figure}
\includegraphics[width=8cm]{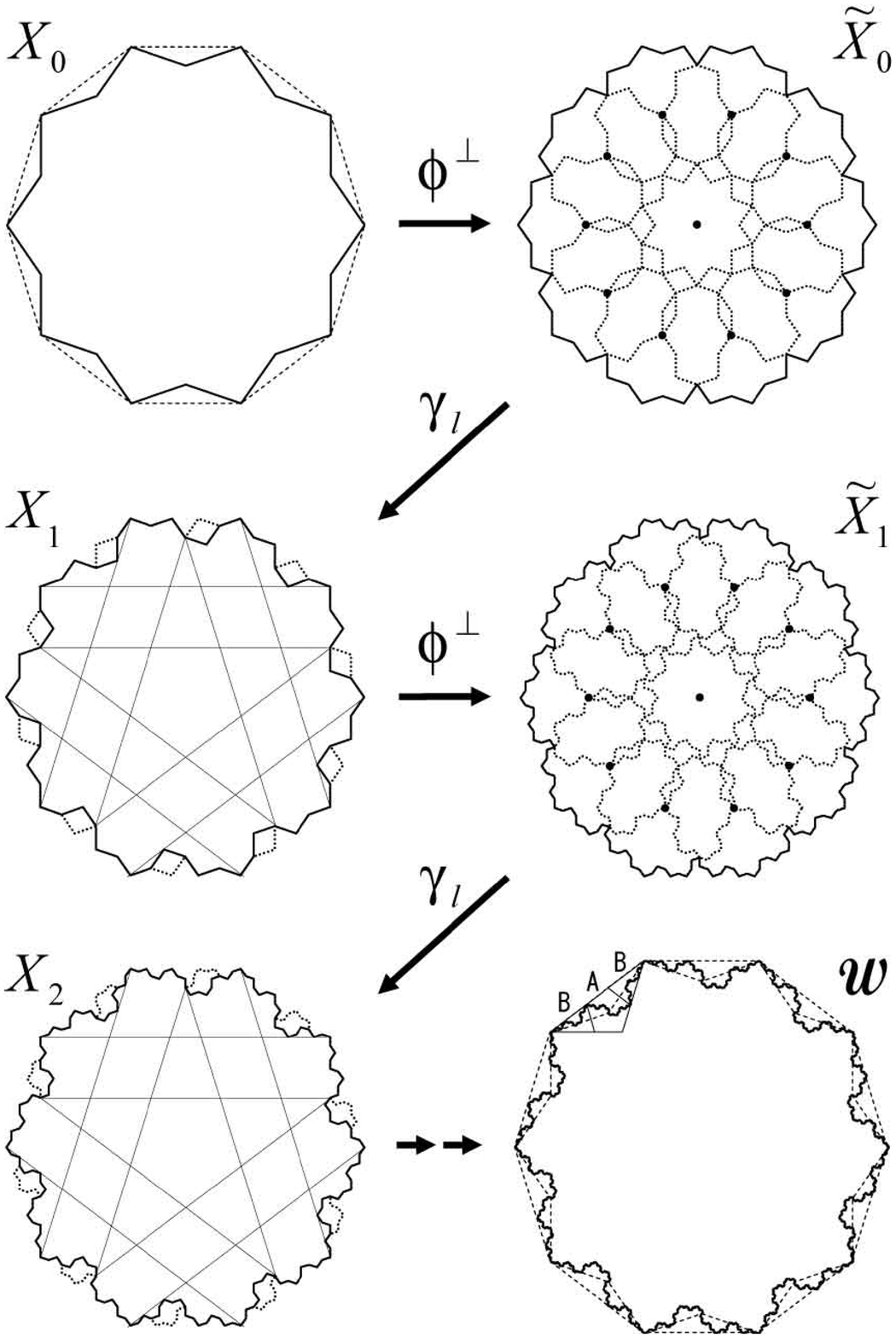}
\caption{The dual map for the left handed GPP is illustrated.
Starting from a star decagon $X_0$ for a non-chiral RPH tiling,
the dual set map $\phi^\perp$ and the subtraction process $\gamma_l$
is alternately applied. For applying the dual set map, a copy of
the reduced figure, e.g. $\sigma^\ast X_0$, is placed on every point
of $\mathcal{S}^\perp$ indicated by the black dots. For $\gamma_l$,
a single
end of all the ten strips are carved as described in the text.
Only five strips are depicted by the thin lines. By iteration, the
composite of the two maps, $\gamma_l\cdot\phi^\perp$, generates a
series of figures $X_i$ ($i=0,1,2,...$) (thicker solid lines),
each of which corresponds to a ternary RPH tiling. The limit figure
$X_\infty$ is the boundary of the atomic surface $\mathcal{W}$
for the RPH tiling generated solely by the left-handed GPP. Note
that a triangular region close to the boundary of $\mathcal{W}$
is divided into a central regular pentagon (A), $3/5$ of which is
occupied, and the two adjacent isosceles triangles (B), half of
which in total is occupied.}
\label{fig8}
\end{figure}

\newpage

\begin{figure}
\includegraphics[width=7.5cm]{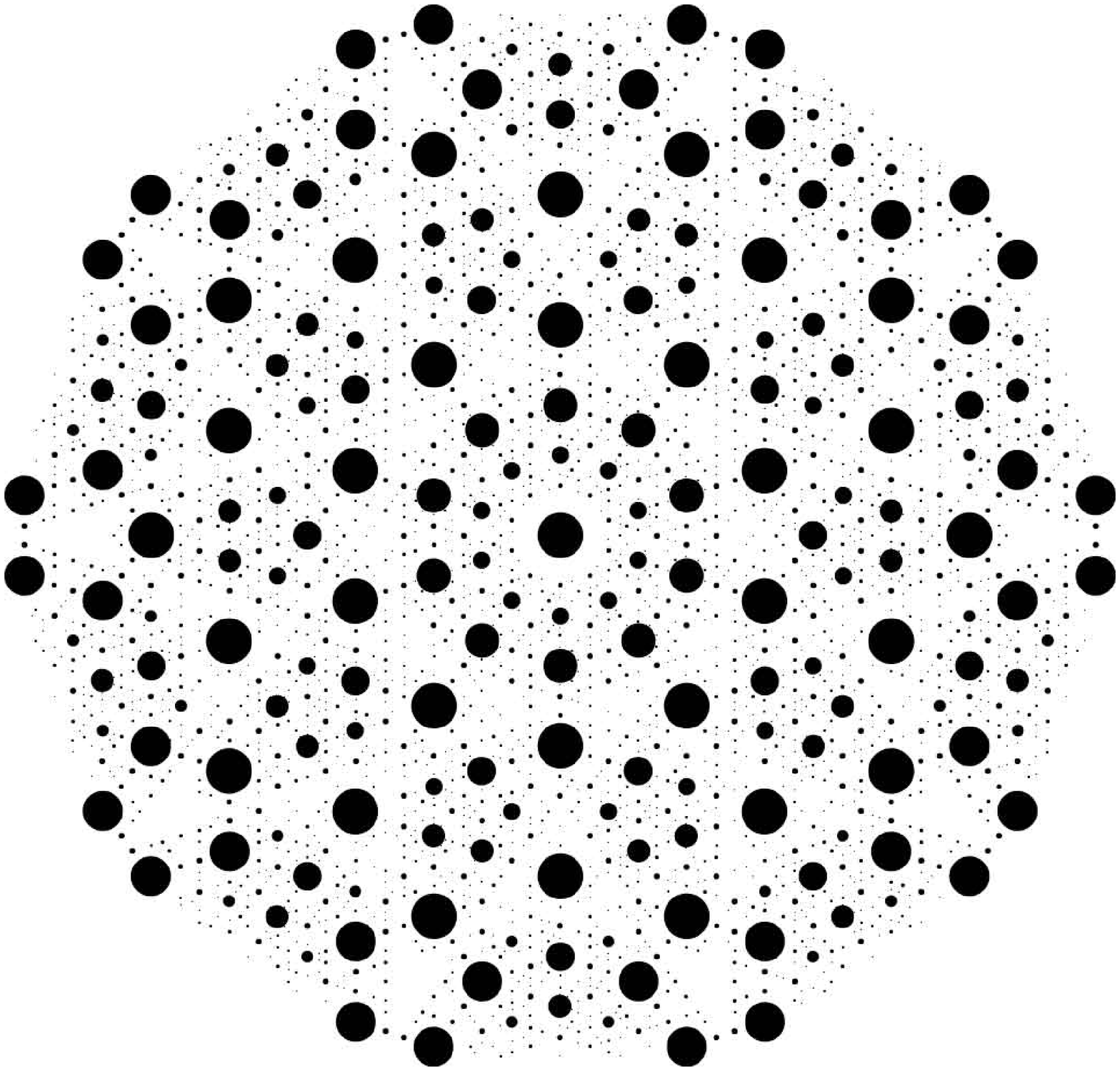}
\caption{The structure factor of the RPH-tiling generated by
the left-handed GPP depicted in Fig.\ref{fig4}. The area of
each spot corresponds to the intensity. The chirality is
manifested in the weakest spots.}
\label{fig9}
\end{figure}



\end{document}